\documentclass[english,twocolumn,a4]{revtex4}
\usepackage[T1]{fontenc}
\usepackage[latin9]{inputenc}
\usepackage{graphicx}

\makeatletter

\usepackage{babel}
\makeatother

\begin{document}
\title{Theoretical analysis of oxygen vacancies in layered sodium cobaltate
Na$_{x}$CoO$_{2-\delta}$}

\author{Simone Casolo$^{a}$}

\email{simone.casolo@unimi.it}

\author{
Ole Martin L\o vvik$^{b,c}$
}
\author{Harald Fjeld$^{d}$}
\author{Truls Norby$^{d}$}
\affiliation{
$^a$ Dipartimento di Chimica,
Universit\'a degli Studi di Milano, via Golgi 19, 20133 Milan, Italy.\\
$^b$ Department of Physics, University of Oslo, P.O. Box 1048 Blindern, 
NO-0316 Oslo, Norway\\
$^c$ SINTEF, Materials and Chemistry, Forskningsvn. NO-0314 Oslo, Norway\\
$^d$ Department of Chemistry, University of Oslo, Centre for Materials Science and Nanotechnology, FERMiO, Gaustadalleen 21, NO-0349 Oslo, Norway\\
}

\begin{abstract}
Sodium cobaltate with high Na content is a promising thermoelectric 
material. 
It has recently been reported that oxygen vacancies can alter the material 
properties, reducing its figure of merit. However, experimental 
data about the oxygen stoichiometry are contradictory.
We therefore studied the formation of oxygen vacancies in Na$_{x}$CoO$_2$ 
with first principles calculations, focusing on $x$=0.75. 
We show that a very low oxygen vacancy concentration 
is expected at the temperatures and partial pressures relevant 
for applications. 

\end{abstract}

\maketitle

\section{Introduction}

Layered sodium cobaltate Na$_x$CoO$_2$ ($x\leq$ 1) 
has in recent years attracted 
considerable attention because of its remarkable electronic and magnetic 
properties. Most of the interesting physical properties of this material 
come from its quasi 2-dimensional CoO$_2$ layers where cobalt 
exists as Co$^{3+}$ and Co$^{4+}$ ions, surrounded by intercalating Na$^+$ 
ions \cite{Hinuma,Zandbergen,MengJCP}. 
Understanding how the oxide layer is affected by temperature and sample 
composition is therefore crucial when interpreting the physical properties.\\
By changing the sodium content, one can continuously tune the
oxidation state of the cobalt ions.
In NaCoO$_2$ ($x=1$) all cobalt atoms are found in a Co$^{3+}$ state 
and the material is an insulator, whereas for 
$x$ < 1, negatively charged sodium vacancies are introduced.  
Na$_x$CoO$_2$ is known to be a $p$-type conductor with metallic 
conduction behavior in which itinerant electron holes give rise to  
a particularly high thermoelectric effect\cite{Terasaki97,Ong03}. \\
The influence of point defects other than Na vacancies on   
sodium cobaltate is still debated.  
Their concentrations are linked through the electroneutrality requirement 
and will thus vary depending on Na content, temperature and oxygen activity
giving rise to a complex scenario.
Among the possible defects, it has been recently reported that oxygen vacancies 
could have an interesting two-fold effect on the 
thermoelectric properties in this material. 
Firstly, they can affect the cobalt oxidation state, 
\emph{i.e.} reducing the charge carrier concentration and 
increasing the thermopower.
Unfortunately, such defects also modify phonon scattering so that the 
overall effect is a reduction of the figure of merit of 
Na$_{x}$CoO$_{2-\delta}$ with increasing the defect
concentration ($\delta$) \cite{truls2010}.\\ 
Experimental data about the concentration of vacancies in 
sodium cobaltate are contradictory, especially at $x \sim 0.75$.  
While some groups have identified an oxygen non-stoichiometry 
up to $\delta$ = 0.16 
\cite{Shu10,MechSpecNaCoO,stoklosa,truls2010,banobrelopez,chou2005}, 
others found no vacancies within the experimental 
error \cite{Alloul11,viciu,choi2006}. 
In order to clarify this, we have in this work studied
the oxygen vacancy formation in Na$_{0.75}$CoO$_{2}$ by first principles 
density functional theory (DFT) calculations. From these results we
computed free energies of formation, at various temperatures and 
oxygen partial pressures. We found that oxygen vacancies have high 
free energies of formation even at high temperature, suggesting that the 
experimentally observed weight losses attributed to oxygen vacancies
 may have a different origin.\\  

\section{Theoretical Model}

We assumed the following reaction (using Kr\"oger-Vink notation): 
\begin{equation}
\mbox{Na$_x$CoO$_2$} \rightleftharpoons \mbox{Na$_x$CoO$_{2-\delta}$} 
+ \delta \mbox{V}_{\mbox{o}}^{q\bullet} + q\delta e' + \frac{\delta}{2}\mbox{O}_2
\end{equation}
The Gibbs' free energies of formation 
for oxygen vacancies $\Delta G_f(P,T)$ at a pressure $P$ and temperature $T$ were accordingly computed from\cite{Northup}
\begin{equation}\label{eq1}
\Delta G_f(P,T) = G_{def}(P,T) - G_{perf}(P,T) + \frac{1}{2}\mu_{O_2}(P,T) 
+ q\mu_e
\end{equation}
In Eq.\ref{eq1} $G_{def}(P,T)$ and $ G_{perf}(P,T)$ are the Gibbs' 
free energy of the defective and pristine supercells respectively, 
$q$ is the charge state per defect, and 
$\mu_i$ is the chemical potential of the species $i$.\\
Thermal contributions of the solid were considered as 
negligible ($G_{i}(P,T)= H_{i}(P^0,0)$) 
with respect to those of the oxygen gas, whose chemical 
potential $\mu_{O_2}(P,T)$ was calculated from
\begin{equation}\label{eq2}
\mu_{O_2}(P,T) = H_{O_2}(P^0,0) + \Delta \mu_{O_2}(P^0,T)+ k_BT 
\ln \left ( \frac{P}{P^0} \right ),
\end{equation}
where $k_B$ is the Boltzmann constant and $\Delta\mu_{O_2}$ 
is the change in the chemical potential when moving
from $T$=0 to $T=T$ at constant pressure $P^0$= 1 bar,
taken from thermodynamic tables. 
Enthalpies at standard pressure and T=0 K, 
$H_{i}(P^0,0)$, were taken as the calculated DFT total energies and
the  corresponding concentrations of defects were then computed following:
\begin{equation}
\delta = n \times e^{-\frac{\Delta G_f(P,T)}{k_BT}} 
\end{equation}
where $n$ is the number of oxygen atoms per formula unit.\\
The system charge $q$ was simulated by adjusting the total number of electrons 
in the supercell and at the same time adding a compensating jellium 
background to avoid diverging Coulomb contributions.
The contribution of the background charge 
to the defect formation energy was taken into account by 
a correction term $\gamma$, here considered as the shift in the average 
electrostatic potentials at a bulk-like lattice site far from the vacancy 
in the defective ($V_{def}$) and pristine ($V_{perf}$) supercell\cite{valle} 
\begin{equation}
\gamma = V_{def}-V_{perf}
\end{equation} 
Then, this correction was included in the 
electron chemical potential, $\mu_e$ determined by the Na$_x$CoO$_2$ 
Fermi energy $E_F$ \cite{Mattila}.
\begin{equation}\label{eq3}
\mu_e = E_F+\gamma
\end{equation}\\
Periodic DFT as implemented in the VASP package \cite{VASP1} 
was used throughout this work.
A spin polarized gradient corrected Perdew-Burke-Ernzerhof (PBE)
functional \cite{PBE1} was used with a plane wave energy cutoff of 600 eV.
Core electrons were included through the projector augmented wave (PAW) 
method \cite{PAW1}.
Where needed we included an on-site Coulomb interaction (GGA+$U$) 
in order to localize cobalt $d$ electrons\cite{LDA+U}. The parameters for 
this system ($U$=5.0 eV and $J$=0.965 eV) were taken from ref.\cite{Okabe} 
and are very similar to those used in other studies of the electronic structure 
of CoO$_2$-based oxides \cite{Louie,Hinuma,Zou}.\\
The reciprocal space was sampled by a $\Gamma$-centered $\bf k$-point grid 
in which the maximum distance between points is 0.15 $\times 2 \pi$/$|a|$.
Ionic positions were relaxed until the maximum
force was lower than 0.03 eV\AA$^{-1}$. Simultaneous relaxation of the
lattice parameters was performed for the defect free structures.\\
To sample the many possible sodium ordered structures 
we adopted three different models for Na$_{0.75}$CoO$_{2}$. They are
shown in Fig.\ \ref{fig:struc} and named 
 \emph{diamond, filled honeycomb (FHC)} and \emph{zigzag}. 
They are based on a tetragonal $\sqrt{3}\times 4\times 1$ supercell
consisting of 16 formula units \cite{Balsys} and  they 
represent different relative concentrations 
of the two inequivalent Na crystallographic sites, Na1  
and Na2 (Na1/Na2=1, 1/2, and 1/5). 
The three models are based on the experimental structure reported
by Zandbergen \emph{et al.}\cite{Zandbergen} 
(diamond), and on
the theoretical study of Meng \emph{et al.}\cite{Meng2005,MengJCP}
(FHC and zigzag).
These were chosen as a representative selection of simple 
periodic models from 
the large manifold of structures, suitable for a first principles study.
Nevertheless, Na self-diffusion is an efficient process already at 
room temperature \cite{Na_melt},
so we expect a strong Na disorder at high $T$.\\ 

\section{Results and Discussion}

\begin{figure}[!t]
\begin{centering}
\includegraphics[clip,width=0.99\columnwidth]{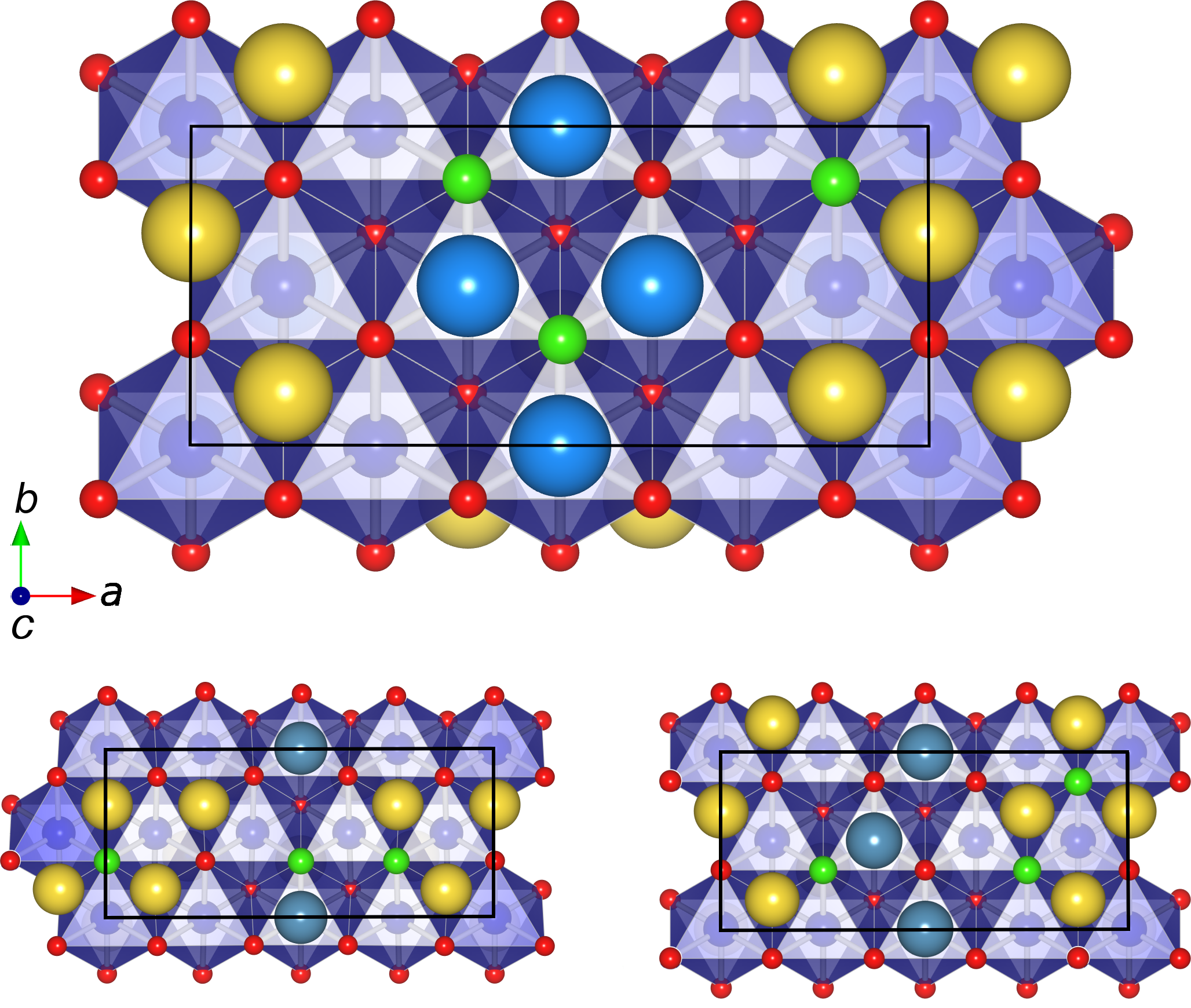}
\par\end{centering}
\caption{Structure of the three Na$_{0.75}$CoO$_2$ models, in their
starting (non relaxed) positions, as looking along the 00$\bar{1}$ 
direction.
Top panel: diamond, Bottom left panel: FHC, Bottom right panel: zigzag
model. Na1 sites are shown as light blue balls, and Na2 as yellow balls.
Co ions are shown as dark balls at the center of the blue CoO$_6$ octahedra.
All the oxygen sites are shown in red, but the ones removed in the vacancy
calculations, shown in green.
\label{fig:struc}}
\end{figure}
The three 
supercells used as staring points for creating oxygen vacancies are 
shown in Figure \ref{fig:struc}. The corresponding lattice parameters
of our relaxed 
models ($a$=11.45, $b$=4.99, $c$=10.88~\AA) and the buckling of the 
CoO$_2$ layer ($\pm$0.05 $c$)\cite{Roger} are in good agreement 
with those determined by neutron diffraction experiments
($a$=11.37, $b$=4.92, $c$=10.81~\AA)\cite{Huang}.
\begin{figure*}[!t]
\begin{centering}
\includegraphics[clip,width=1.6\columnwidth]{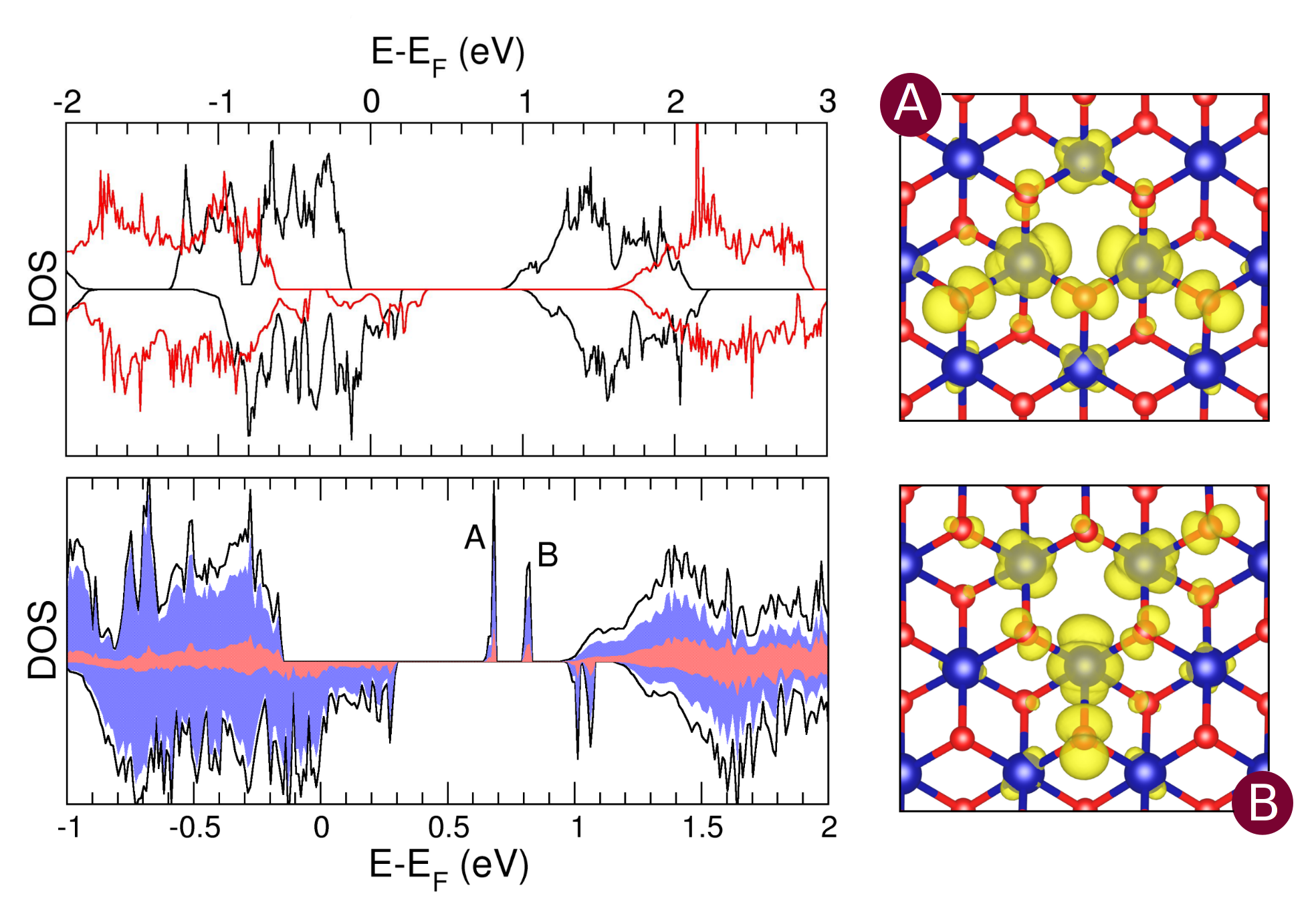}
\par\end{centering}
\caption{
  Top panel: Density of states for Na$_x$CoO$_{2}$ computed with GGA 
  (black line) and GGA+$U$ (red line). Lower panel: Density of states
  for Na$_x$CoO$_{2-\delta}$ computed with GGA. Spin up and down component are
  shown with positive and negative values respectively. 
  Cobalt $d$ and oxygen $p$ projected density of states are shaded in blue 
  and red respectively for the diamond structure (see text). 
  Charge density isosurfaces of the two localized states in the gap (A and B), 
  are shown in right the panel.
\label{fig:electro}}
\end{figure*}
Oxygen vacancies were considered in their 
neutral, +1 and +2 charge states (V$_{\mbox{o}}^{\times}$, V$_{\mbox{o}}^{\bullet}$ 
and  V$_{\mbox{o}}^{\bullet\bullet}$ in Kr\"oger-Vink notation), created at 
three different positions for each of the structural models. 
Oxygen atoms were removed from the sites shown 
as green balls in Fig.\ref{fig:struc}, then the whole lattice structure
was relaxed to its equilibrium geometry.
The electronic structure (density of states, DOS) of the perfect and 
defective Na$_{0.75}$CoO$_{2}$ (diamond structure) is shown in Figure 
\ref{fig:electro}. 
Firstly we note that the electronic structures we computed are half-metallic,
both for perfect and defective lattices, and independent of 
using GGA or GGA+$U$, in agreement with similar studies\cite{Louie,Zou}.   
The various sodium ion orderings and vacancy positions we have considered
did not give any qualitative or quantitative differences in the DOS. 
Indeed Na states are found only at much more negative energies in the 
valence 
band, confirming that the intercalating ions simply provide charge carriers
to the CoO$_2$ layers. \\
For the pristine structure the dominant contributions to the DOS 
close to the Fermi level come from 
cobalt $d$ and oxygen $p$ states in the oxide layer, split by the distorted 
octahedral crystal field into $t_{2g}=e'_{g} \oplus a_{1g}$ 
manifolds\cite{Lepetit1}. 
Applying Bader charge analysis to our GGA results 
we found that all the Co ions in the 
pristine structures are neither in a +3 or +4 state, but rather 
that holes are delocalized on all the transition metal atoms.
When the $U$ term is considered, the holes are instead localized on 
few Co ions, which $d$ states are pushed to lower energies into the 
valence band, so that oxygen $p$ states are now dominant at the 
Fermi energy (Fig.\ref{fig:electro}). 
\begin{figure}[!t]
\begin{centering}
\includegraphics[clip,width=0.99\columnwidth]{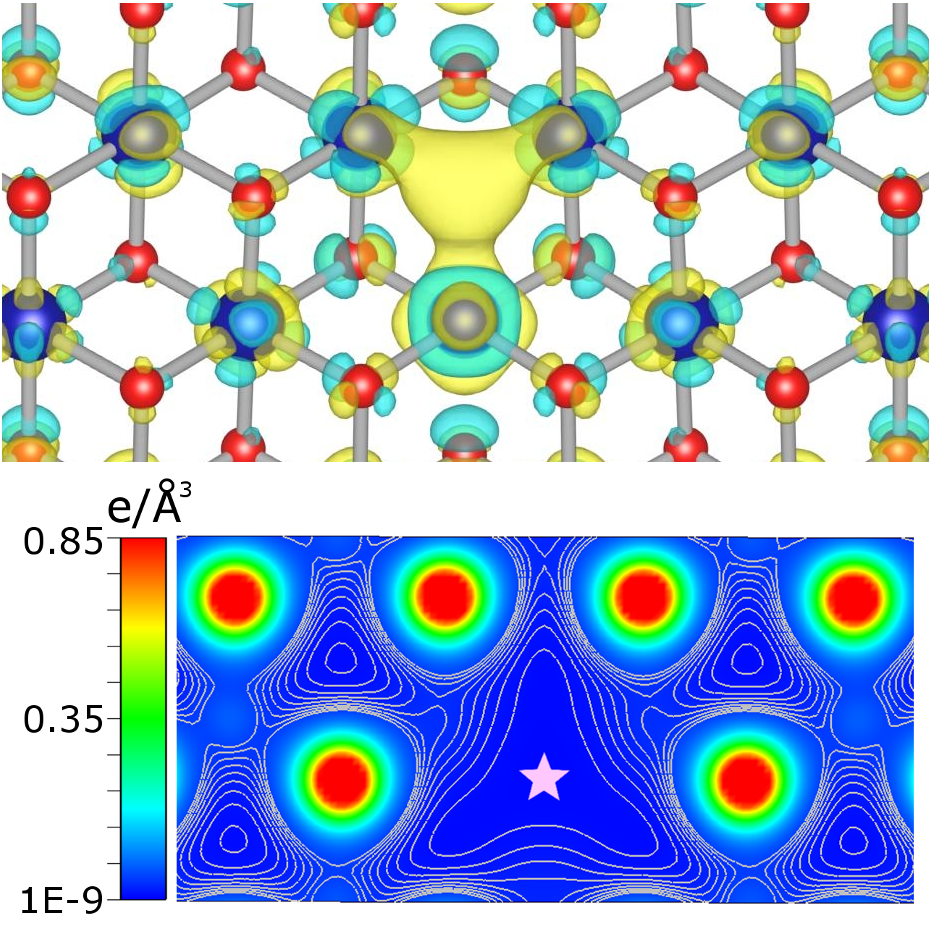}
\par\end{centering}
\caption{Upper panel: Difference of electron densities for 
  V$^{\times}_{\mbox{o}}$ and V$^{\bullet\bullet}_{\mbox{o}}$ defects
in $diam1$ structure  as seen from the 00$\bar{1}$ direction. 
  Atoms color code follows that of Figure 1. 
  Isosurfaces show the value +0.05 $e$ (yellow) and -0.05 $e$ (light blue).
  Lower panel: Charge density in the plane parallel to 
  (001) containing the V$_{\mbox{o}}^{\bullet\bullet}$ defect 
  (site shown with a star) in Na$_{0.75}$CoO$_{2-\delta}$. 
\label{fig:chgcar}}
\end{figure}
However, the Hubbard-like $U$ term is an arbitrary parameter 
used to correct GGA by applying an energy penalty to double 
occupation of $d$ orbitals, 
for which many different values have been proposed ranging from 
4.0 eV \cite{wang} to 5.5 eV\cite{Louie}.
While GGA+$U$ was proven to localize holes in this class of materials
it is not known how reliable the ground state energies 
(and their differences) are, in particular for metallic or half-metallic 
systems. Recently, Hinuma 
\emph{et al.} showed that GGA+$U$ performs worse than GGA in 
reproducing experimental data for Na$_x$CoO$_2$  when $x>$0.6, 
in particular its lattice parameters and the formation energy of different
Na-ordered ground states \cite{Hinuma}. 
Moreover, we note that 
within our approach the free energy of formation (Eq.\ref{eq1} and \ref{eq3}) 
is dependent on the position of the Fermi level of Na$_{0.75}$CoO$_2$.
By introduction of the Hubbard-like term, 
the valence band edge (hence the Fermi 
level) shifts to lower energies, making the calculated free energy 
of formation strongly dependent on the arbitrarily chosen value of $U$.
In GGA, the homogeneous oxidation state of Co ions
 is due to the self-interaction error that 
tends to over-delocalize the electron density. Still this  scenario is
compatible with recent NMR measurements which predict holes 
delocalization in Na rich cobaltaltes rather than localized Co$^{+4}$ 
sites\cite{Julien08,Lang}. It is also known that at low temperature 
different Na orderings establish 
patterns of Co$^{3+}$/Co$^{4+}$ 
which may modulate the formation energies of vacancies in 
neighbouring oxygen sites, and this effect would not be correctly 
represented 
within our computational approach. Nevertheless, at high temperature
the sodium sublattice is liquid-like\cite{Na_melt} and fast diffusing Na ions 
would average out this effect. 
We then expect our GGA approach to be appropriate for
representing qualitatively the high $T$
phase, while it may fail 
representing the fine modulation induced by the spin states of 
cobalt ions at low temperature. Therefore, in this work we chose to 
rely only on free energies of formation computed with GGA, that are free 
from any arbitrary term.\\
By looking at the GGA electronic structure of the defective system  
in Figure \ref{fig:electro} we notice an enlargement of the band gap (of
about 25 $\%$ for the spin majority component), 
in which two sharp defect states lying at about 0.12 eV 
from each other are found. 
Charge density isosurfaces corresponding to the defect states are very 
similar in shape, and mostly localized on the vacancy nearest neighbors. 
Using a simple tight-binding argument for the bipartite CoO$_2$ lattice it is 
possible to predict how the removal of two $p$ orbitals at the oxygen site, 
i.e. upon vacancy formation, as many degenerate states 
(known as \emph{midgap states}) localize on the Co 
sublattice\cite{TheoInui,TheoLieb}.
In this case we suggest 
that the difference in symmetry and energy of the two defect levels may be 
caused by the Coulomb potential generated by the Na ions, which modulates 
the on-site energies of Co and O sites\cite{Marianetti}.\\
The analysis of the ground state electronic structure 
revealed that in the neutral defect calculation
V$_{\mbox{o}}^{\times}$, electron density accumulates at the vacancy site and 
on the three nearest neighbouring Co ions. 
The difference in electron
density for the neutral and doubly charged system is shown in 
Figure \ref{fig:chgcar}.
This difference in charge is too small 
to give a perfectly neutral defect site, but rather a partially positively 
charged vacancy surrounded by three Co ions which are 
reduced by 0.2 electrons each. 
We note here that GGA+U gives qualitatively the same results.   
The charge density at the vacancy site is removed by progressively
increasing the system charge, resulting in an effectively doubly 
positively charged oxygen vacancy V$_{\mbox{o}}^{\bullet\bullet}$, as expected.
The same analysis showed that Co neighbors
have been reduced by 0.05 $e$/atom when increasing the system charge
to $q=2$.\\
Gibbs' free energies of formation ($\Delta G_f$) for oxygen vacancies 
are shown in Table I
for the high temperature regime ($T=1000$ K) relevant for applications as 
thermoelectric material. 
Results are reported for the three different models in
both a $\sqrt{3}\times 4\times 1$ and $2\sqrt{3}\times 4\times 1$ supercell,
corresponding to Na$_{0.75}$CoO$_{2-\delta}$ with 
$\delta=0.031$ and $\delta=0.016$, respectively.
The table values were based on the Fermi energy $E_F$ of the perfect 
material and $P(O_2)=1.0$ atm.\\
\begin{table}[!t]\label{tab:energies}
\begin{tabular}{lcccccccc}
\hline \hline
 & &  \multicolumn{7}{c}{$\Delta G_f$ (eV)}\\
 & & \multicolumn{3}{c}{$\delta$=0.031} & &  \multicolumn{3}{c}{$\delta$=0.016}\\
\hline
   & & V$_{\mbox{o}}^{\bullet\bullet}$ & V$_{\mbox{o}}^{\bullet}$ &  V$_{\mbox{o}}^{\times}$ &  &
  V$_{\mbox{o}}^{\bullet\bullet}$ & V$_{\mbox{o}}^{\bullet}$ &  V$_{\mbox{o}}^{\times}$ \\
  \hline
 diam1   & $\qquad$ &  2.19 & 2.26 & 2.30 & $\qquad $ & 2.27 & 2.27 & 2.28 \\ 
 diam2   & &  2.31 & 2.34 & 2.36 & & 2.32 & 2.34 & 2.36 \\ 
 diam3   & &  2.29 & 2.33 & 2.36 & & 2.35 & 2.37 & 2.39 \\
 zigzag1 & &  2.35 & 2.37 & 2.37 & & 2.36 & 2.37 & 2.39 \\
 zigzag2 & &  2.32 & 2.37 & 2.39 & & 2.39 & 2.40 & 2.40 \\
 zigzag3 & &  2.37 & 2.42 & 2.44 & & 2.41 & 2.43 & 2.45 \\
 FHC1    & &  2.46 & 2.44 & 2.48 & & 2.40 & 2.39 & 2.42\\
 FHC2    & &  2.37 & 2.39 & 2.41 & & 2.38 & 2.37 & 2.39 \\
 FHC3    & &  2.19 & 2.24 & 2.26 & & 2.20 & 2.23 & 2.25 \\
\hline\hline
\end{tabular}
\caption{Defect formation free energy as defined in Eq.\ \ref{eq1} 
for neutral, singly and doubly charged oxygen 
vacancy in  Na$_{0.75}$CoO$_{2-\delta}$ computed for two different $\delta$ 
values. $T=1000$ K and $P(O_2)=1.0$ atm. All the values are in eV.}
\end{table}
Overall, the defect formation energies shown in Table I 
are quite high, ranging from 2.19 to 2.48 eV at $T=1000$ K and
$P(O_2)=1.0$ atm, implying a very low concentration of vacancies.
Formation energies are distributed in a narrow range of 0.2~eV, with 
no clear preference for any of the structural models. 
\begin{figure*}[!t]
\begin{centering}
\includegraphics[clip,width=1.40\columnwidth]{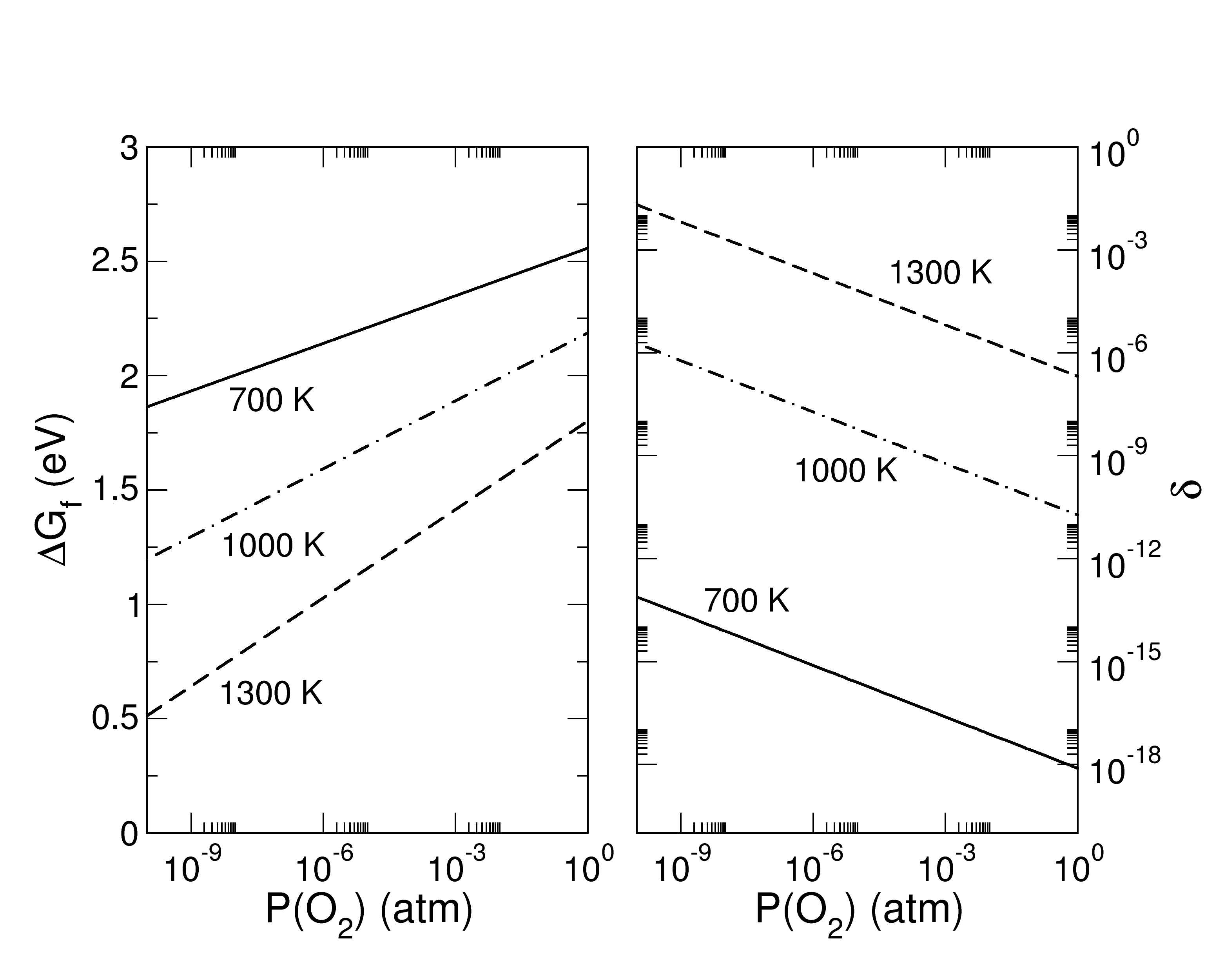}
\par\end{centering}
\caption{Formation free energy (left) and concentration (right) of
doubly charged oxygen vacancies in the $diam1$ structure
as a function of oxygen partial pressure.
Three different temperatures are investigated.
\label{fig:pressure}}
\end{figure*}
This suggests that the Na 
ordering only plays a minor role for the vacancy formation 
energy. As well, the difference in the $\Delta G_f$ for neutral and charged 
defects is also small at both the concentrations $\delta$ considered, 
and lower than 0.1~eV in most of the cases.
This suggests that the Coulomb potential
generated by the charged defect is efficiently screened already at the 
length scale of the shortest lattice parameter, {\em i.e.} $\sim$5.0 \AA.
Nevertheless, it appears that the doubly charged defect
V$_{\mbox{o}}^{\bullet\bullet}$ is
slightly favored compared to the other two possible charges, suggesting a 
non-perfect dielectric screening. This small energy difference is a result of
Na$_{0.75}$CoO$_{2-\delta}$ being a $p$-type conductor, 
with unfilled valence states.
These states lay at energies only slightly above the Fermi level 
(see Fig.\ref{fig:electro}), and are readily filled by the extra electron(s)
accompanying the singly charged and neutral oxygen vacancies.\\
In order to quantify the concentration of vacancies under relevant
experimental conditions we calculated $\Delta G_f$ for $x=0.75$ at 
different temperatures for the structure with the lowest free energy of 
formation (2.19 eV) in Table 1. 
Results are shown in 
the left panel of Figure \ref{fig:pressure} for three 
different temperatures (700, 1000 and 1300 K) as a function of 
$P({\rm O}_2)$.
In the right panel we show the corresponding concentration of vacancies per 
formula unit, $\delta$. We also note that
our formation energies are in good agreement with those recently
reported by Yoshiya {\em et al.} \cite{Fisher2010} for $x=0.50$ and $x=1$.
It is clear that the oxygen vacancy free energy of formation is very high 
at relevant temperatures and oxygen partial pressures. 
Even at extreme experimental conditions 
such as $T=1300$ K (close to the homogeneous melting point\cite{Peng2009}) 
and $P(O_2)=10^{-7}$ atm the vacancy concentration would be as 
little as $\delta \simeq 6.5\times10^{-4}$. 
If more common experimental conditions are chosen, {\em e.g.} $T=700$ K and
$P(O_2)=10^{-1}$ atm, we obtain $\delta\sim 10^{-18}$. This is several 
orders of magnitude lower than reported in some previous publications, 
{\em e.g.} Ref.\ \cite{truls2010}.\\
Having ruled out the Na ordering as the main influence on the vacancy 
formation 
energy we now test the role of sodium vacancies, thus indirectly that of
the Co oxidation state. 
To do this we modify the sodium content $x$, as it is known 
to determine directly the magnetic behaviour 
(hence the oxidation state) 
of Co ions by supplying electrons to the oxide layers\cite{Lang}. 
We therefore studied the V$^{\bullet\bullet}_{\mbox{o}}$ 
defect formation for a range of Na concentrations, in order to 
span as much as 
possible the possible cobalt charges, from +4 ($x\rightarrow$ 0) 
to +3 ($x\rightarrow$ 1).
Two different Na orderings were generated for each $x$ value starting from the 
diamond structure,
and the oxygen vacancy was created at three different sites in 
each of those. 
Averaged GGA results for all the structures considered 
are shown in Fig.\ref{fig:comp}, using the same 
$T$ and $P$(O$_2$) conditions as 
in Table 1. 
\begin{figure}[!t]
\begin{centering}
\includegraphics[clip,width=0.80\columnwidth]{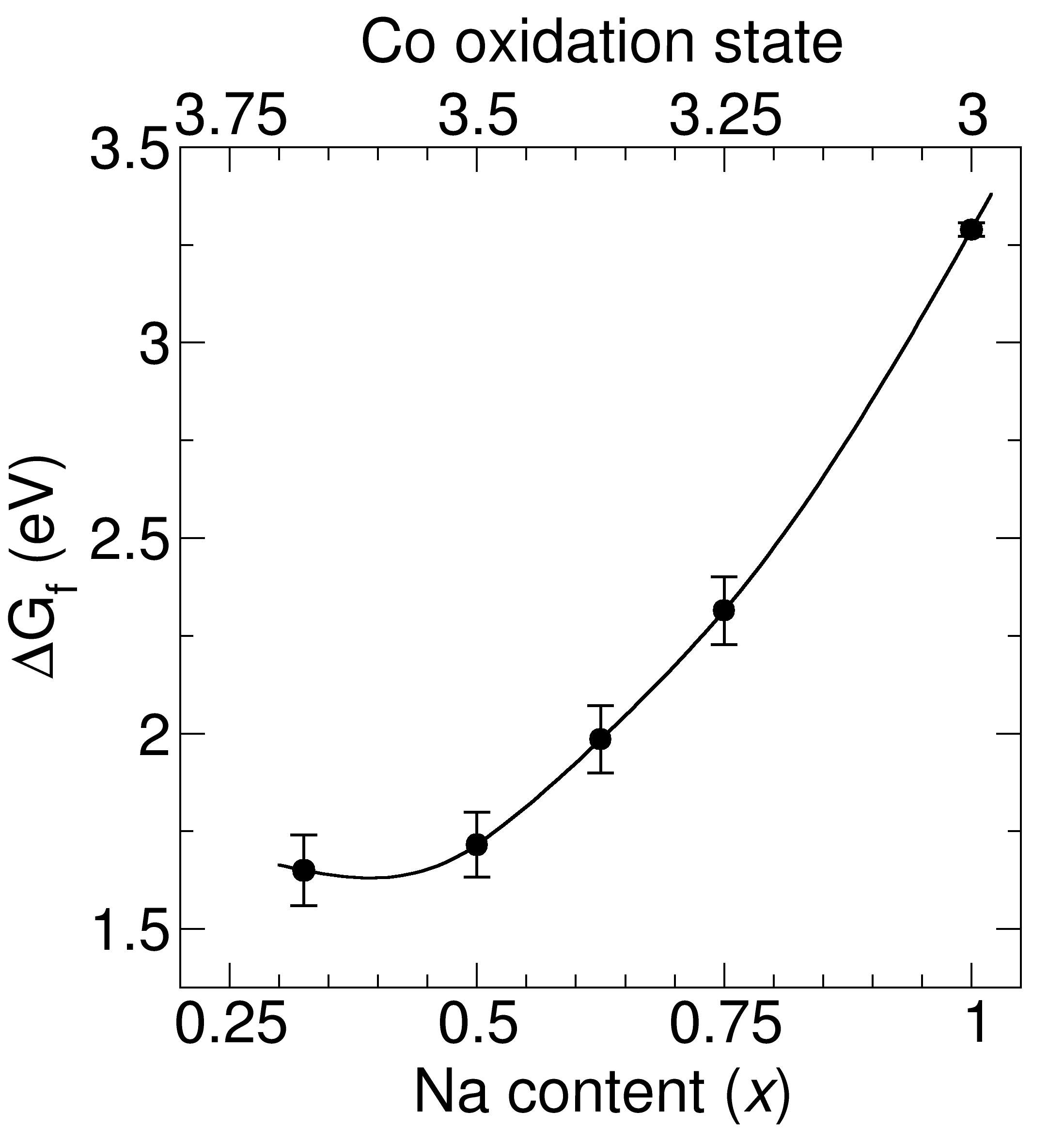}
\par\end{centering}
\caption{Formation energy for the doubly charged oxygen vacancy as a
function of the Na content, \emph{i.e.} of the Co oxidation state.
Each formation energy has been calculated from a few
structures with different ordering, mean values and standard deviations
are shown. The line is a guidance for the eye. See text for details.\label{fig:comp}}
\end{figure}
For Na rich structures, the change in the formation energies with 
$x$ is substantial, but the absolute energies are still very large. 
When the Na content is reduced, $\Delta G_f$ lowers almost linearly down to 
$x$=0.50, where it reaches a plateau of about 1.70 eV, suggesting
that Co$^{4+}$ ions favor the formation of O vacancies. 
However, the magnitude of the free energy of formation 
is still high, so that even at low sodium content ($x\leq$0.5) 
the concentration of oxygen vacancies under relevant experimental 
conditions is expected to be very low.
A detailed analysis of the effect of oxygen vacancies on the Co oxidation 
state or \emph{vice versa} goes beyond the scope of this study.\\

\section{Conclusions}

In conclusion, we have shown that the equilibrium concentration 
of oxygen vacancies in sodium cobaltate is very low at conditions 
relevant for applications
as thermoelectric material.
Our results may help to settle the disagreement 
between experimental reports on the oxygen 
stoichiometry in Na$_{x}$CoO$_{2}$. At high sodium content, 
concentrations of vacancies ranging from 
$\delta$ = 0.05 to 0.16 have been 
reported\cite{Shu10,banobrelopez,chou2005,truls2010}, but 
other studies have concluded that the oxygen concentration is 
indeed stoichiometric\cite{viciu,choi2006,Alloul11}. 
Our results are clearly consistent with the latter ones.\\
We can here only speculate on the reason for this discrepancy. 
One possible explanation is existence of secondary phases in some of the
experiments, where these phases are responsible for the oxygen weight
loss upon reduction of oxygen partial pressure. 
Other possibilities comprise Na evaporation, 
the presence of other defects than considered here, 
or unintended interaction between the sample and experimental setup.\\

\section{Acknowledgments}

We acknowledge Tor Svendsen Bj\o rheim for
helpful suggestions, the Research Council of Norway 
(Renergi project THERMEL-143386) for economic support, 
and the NOTUR consortium for providing access to their computational 
facilities. \\


\end{document}